\documentclass[11p]{article}
\newcommand{\be}{\begin{equation}

gin{equation}}
\newcommand{\ee}{\end{equation}}
\newcommand{\bea}{\begin{eqnarray}}
\newcommand{\eea}{\end{eqnarray}}

\begin{document}
\title{A note on wealth in a volatile economy}
\author{Matteo Marsili\\
Abdus Salam International Centre for Theoretical Physics\\
Strada Costiera 11, 34014 Trieste, Italy}
\maketitle

\begin{abstract}
  I show that if the capital accumulation dynamics is stochastic a new
  term, in addition to that given by accounting prices, has to be
  introduced in order to derive a correct estimate of the genuine
  wealth of an economy. In a simple model with multiplicative
  accumulation dynamics I show that: 1) the value function is always a
  decreasing function of volatility 2) the accounting prices are
  affected by volatility 3) the new term always gives a negative
  contribution to wealth changes. I discuss results for models with
  constant elasticity utility functions. When the elasticity of
  marginal utility is larger than one, accounting prices increase with
  volatility whereas when it is less than one accounting prices
  decrease with volatility. These conclusions are not altered when
  adopting optimal saving rates.
\end{abstract}
\vskip 0.5cm

Estimating genuine wealth for an economy has been the subject of much
recent interest as it is a key instrument to assess sustainability.
In particular Arrow, Dasgupta and Maler \cite{ADM} -- hereafter ADM --
have broadly discussed the issue in a recent paper. In brief, ADM show
that wealth changes over time can be computed from the knowledge of
accounting prices of the various resources in the economy.  Their
approach is mainly deterministic, apart from the case of a discrete
time stochastic model. Here I generalize the treatment to the case of
continuous time stochastic processes and show that the effect of
volatility cannot be expressed in terms of accounting prices. I
discuss a simple model and its application to constant elasticity
utility functions in the next section. I discuss both the case of
imperfect economies and of efficient ones, where the saving rate is
adjusted to its optimal value, as discussed in \cite{LS}.

Let 
\begin{equation}
V(k_0,t_0)=E\int_0^\infty d\tau e^{-\delta\tau}u(c_{t_0+\tau})
\label{V}
\end{equation}
be the wealth function, where $c_{t_0+\tau}$ is the consumption flow
at time $t_0+\tau$. 

We imagine that $k(t)$ follows a stochastic dynamics

\begin{equation}
dk=a(k,t)dt+b(k,t)dW,~~~~~~k(t_0)=k_0
\end{equation}
where $dW$ is the increment of the Wiener process: $E[dW]=0$,
$dW^2=E[dW^2]=dt$. If the process is time homogeneous $a$ and $b$ do
not depend explicitly on $t$. Most results we are going to use are standard
in stochastic processes (see e.g. \cite{CWG}).

Because the situation at time $t_0+\tau$ can no more be
deterministically related to the capital stock $k_0$ at time $t_0$, we
can no more assume that $c_{t_0+\tau}$ is given by a resource
allocation mechanism $\alpha(t_0,t_0+\tau,k_0)$. We assume that
$c_{t_0+\tau}=C[k(t_0+\tau)]$ is a function of capital stock at the
same time, and hence it is itself a random quantity.

Let us consider the change in $V$ in a time interval $dt$. In that
time interval $k$ changes by an amount $dk$ which is random. So

\begin{eqnarray*}
dV(k_0,t_0)&=&E[V(k_0+dk,t_0)]-V(k_0,t_0)\\
&=&\frac{\partial
V(k_0,t_0)}{\partial t_0}dt+\frac{\partial
V(k_0,t_0)}{\partial k_0} E[dk]+\frac{1}{2}\frac{\partial^2
V(k_0,t_0)}{\partial k_0^2}E[dk^2]+\ldots
\end{eqnarray*}
Now, in order to compute $dV/dt$ in the limit $dt\to 0$ we observe
that $E[dk^2]=b^2(k_0,t_0)E[dW^2]+\ldots=b^2(k_0,t_0)dt+\ldots$ and then 

\begin{equation}
\frac{dV}{dt_0}=\frac{\partial
V(k_0,t_0)}{\partial t_0}+\frac{\partial
V(k_0,t_0)}{\partial k_0}a(k,t)+\frac{1}{2}\frac{\partial^2
V(k_0,t_0)}{\partial k_0^2}b^2(k_0,t_0)
\label{ito}
\end{equation}
The first two terms in the r.h.s. are the usual ones, the last --
which we shall call the {\em Ito term} -- is new.

\section{A simple model}

Let us take a simple economy with a single capital good whose stock is
$k(t)$. In a time interval $dt$ the production is

\begin{equation}
dY=(\mu dt+\sigma dW)k
\end{equation}
which means that the production process is subject to stochastic
fluctuations. The resource allocation mechanism is defined by assuming
that in a time interval $dt$ a fraction $\nu dt$ of the capital $k(t)$
is consumed: $dC=\nu k dt$ which gives an utility $dU=u(\nu
k)dt$. The parameter $\nu$ is the consumption rate, which is easily related to the saving rate $1-\nu$, and it may be fixed by economic policy to an optimal value. We shall discuss our results in terms of the consumption rate below, rather than in terms of saving rates.
Capital accumulates according to\footnote{Capital depreciation
can be incorporated into $dC$. More precisely, capital depreciates at
a rate $\gamma$ is equivalent to having $dC=(\nu+\gamma) k dt$, no
depreciation as far as $k$ is concerned. Depreciation can be set back
in by replacing $\nu\to\nu+\gamma$ and $u(x)\to u[(1+\gamma/\nu)x]$ in
the following.}

\begin{equation}
dk=dY-dC=(\mu-\nu)kdt+\sigma k dW,~~~~~t>t_0,~~~~~k(t_0)=k_0
\end{equation}
The wealth function is then given by

\[
V(k_0,t_0)=E\int_{t_0}^\infty  e^{-\delta (t-t_0)}u(\nu k)dt
=\int_{t_0}^\infty  dt e^{-\delta (t-t_0)}\int_0^\infty
P(k,t|k_0,t_0)u(\nu k) dk
\]

Let us compute the distribution $P(k,t|k_0,t_0)$ of $k(t)$ conditional
on $k(t_o)=k_0$. Let us first derive the distribution of $x=\ln k$. We
have

\begin{eqnarray*}
dx=x(t+dt)-x(t)&=&log[1+dk/k(t)]=\frac{dk}{k(t)}-
\frac{1}{2}\left(\frac{dk}{k(t)}\right)^2+\ldots\\
&=&
\left(\mu-\nu-\frac{\sigma^2}{2}\right)dt+\sigma dW
\end{eqnarray*}
where we have used the fact that $dW^2=dt$ and we have neglected all
terms of order higher than $dt$. Notice that the effect of stochastic
fluctuations is to reduce the rate of growth by a term $-\sigma^2/2$.

It is clear that $x(t)$ satisfies a biased random walk:

\begin{equation}
x(t)=\ln k_0+\left(\mu-\nu-\frac{\sigma^2}{2}\right)(t-t_0)+\sigma
W(t)
\end{equation}
where $W(t)$ is a gaussian variable with zero average and variance
$t-t_0$. I introduce, for convenience the parameter 

\begin{equation}
v=\mu-\nu-\frac{\sigma^2}{2}
\end{equation}
which allows us to write

\begin{equation}
V(k_0,t_0)=\int_{0}^\infty  d\tau e^{-\delta \tau}\int_{-\infty}^\infty
\frac{dx}{\sqrt{2\pi\tau}\sigma}\exp\left[-\frac{(x-\ln k_0-v
    \tau)^2}{2\sigma^2 \tau}\right]u\left(\mu e^x\right)
\end{equation}
or, changing variables to $z=\ln k_0+v\tau+\sigma\sqrt{\tau}x$,

\begin{equation}
V(k_0,t_0)=\int_{0}^\infty  d\tau e^{-\delta \tau}\int_{-\infty}^\infty
\frac{dz}{\sqrt{2\pi}}e^{-z^2/2}u\left(\nu
k_0e^{(\mu-\nu-\sigma^2/2)\tau+\sigma\sqrt{\tau}z}\right) 
\end{equation}

\noindent
{\bf Proposition}: If $u"(c)<0$ then $V$ decreases with $\sigma$.

\noindent
{\em Proof:} Let us introduce the shorthand

\begin{equation}
C=\nu k_0e^{(\mu-\nu-\sigma^2/2)\tau+\sigma\sqrt{\tau}z}
\end{equation}
Then
\begin{equation}
\frac{\partial V}{\partial \sigma}=\int_{0}^\infty  d\tau e^{-\delta\tau}
\int_{-\infty}^\infty\frac{dz}{\sqrt{2\pi}}e^{-z^2/2}u'(C)C
[\sqrt{\tau}z-\sigma\tau] 
\end{equation}
If we take a partial integration in $z$ of the term proportional to 
$\sqrt{\tau}z$ we have
\begin{eqnarray*}
\int_{-\infty}^\infty\frac{dz}{\sqrt{2\pi}}e^{-z^2/2}u'(C)C
\sqrt{\tau} z&=&
-\sqrt{\tau}\int_{-\infty}^\infty\frac{dz}{\sqrt{2\pi}}\left[
\frac{\partial}{\partial z}e^{-z^2/2}\right]u'(C)C\\
&=&
-\sqrt{\tau}\left|e^{-z^2/2}u'(C)C\right|_{-\infty}^\infty+
\sqrt{\tau}\int_{-\infty}^\infty\frac{dz}{\sqrt{2\pi}}
e^{-z^2/2}\frac{\partial}{\partial z}u'(C)C\\
&=&
\int_{-\infty}^\infty\frac{dz}{\sqrt{2\pi}}
e^{-z^2/2}[u''(C)C+u'(C)]C\sigma\tau
\end{eqnarray*}
The boundary term in the integration by parts vanishes for any convex utility function $u(C)$. If we combine
this with the previous formula, we get
\begin{equation}
\frac{\partial V}{\partial \sigma}=\sigma\int_{0}^\infty d\tau \tau
e^{-\delta\tau}
\int_{-\infty}^\infty\frac{dz}{\sqrt{2\pi}}e^{-z^2/2}u''(C)C^2
\end{equation}
If $u''(C)<0$, as assumed, then $V$ decreases with $\sigma$.

Note in particular that the time integration kernel $\tau
e^{-\delta\tau}$ has an extra $\tau$ factor which reflects the
fact that uncertainty increases with time. This kernel has a maximal
weight at $\tau=1/\delta$. 

If we use the shorthand $V=\int \ldots u(C)$ then accounting price $p$ is given by
\begin{equation}
k_0 p\equiv k_0\frac{\partial V}{\partial k_0}=\int\ldots u'(C)C.
\label{accp}
\end{equation}
We have seen that
\begin{equation}
\frac{\partial V}{\partial \sigma}=\sigma\int\ldots \tau u''(C)C^2
\end{equation}
then symbolic calculus tells us that 
\begin{equation}
k_0\frac{\partial p}{\partial \sigma}=\sigma\int\ldots \tau
C^2\frac{d^2}{dC^2}u'(C) C=\sigma\int\ldots \tau
C^2[2u''(C)+Cu'''(C)]
\end{equation}
depending on the sign of $2u''(C)+Cu'''(C)$ accounting price will decreases
or increase with $\sigma$.

\noindent
{\bf Proposition} If $u"(c)<0$ then the Ito term is negative.

\noindent
{\em Proof:} By direct calculation, it is easy to see that
\begin{equation}
\frac{\partial^2 V}{\partial k_0^2}=k_0^{-2}\int_{0}^\infty d\tau 
e^{-\delta\tau}
\int_{-\infty}^\infty\frac{dz}{\sqrt{2\pi}}e^{-z^2/2}u''(C)C^2
\label{ito1}
\end{equation}
If $u''(C)<0$, as assumed, then the Ito term gives a {\em negative}
contribution in Eq. (\ref{ito}). 

The change of the value function
for this simple model, form Eq. (\ref{ito}), is given by:

\begin{equation}
\frac{d V}{d t}=
\int_{0}^\infty d\tau e^{-\delta\tau}
\int_{-\infty}^\infty\frac{dz}{\sqrt{2\pi}}e^{-z^2/2}
\left[(\mu-\nu)u'(C)C+\frac{\sigma^2}{2}u''(C)C^2\right]
\end{equation}

Note the similarity of the Ito term $\frac{1}{2}\sigma^2 k_0^2\frac{\partial^2
  V}{\partial k_0^2}$ with 
\[
\sigma^2\frac{\partial V}{\partial \sigma^2}=
\frac{\sigma^2}{2}\int_{0}^\infty d\tau \tau e^{-\delta\tau}
\int_{-\infty}^\infty\frac{dz}{\sqrt{2\pi}}e^{-z^2/2}u''(C)C^2
\]
where it not for the factor $\tau$ in the time integration kernel, it
would be possible to express the Ito term as an accounting price of
volatility.

Finally we remark that optimal consumption rate $\nu^*$ is given by the solution of
\begin{equation}
\nu^*=\frac{\int\ldots u'(C)C}{\int\ldots \tau u'(C)C}=-\left[\frac{\partial \ln p}{\partial \delta}\right]^{-1}.
\end{equation}
The effect of volatility in the case where the optimal consumption rate is implemented can be appreciated from the equation
\begin{equation}
\frac{dV}{d\sigma}=\frac{\partial V}{\partial \sigma}+\left.\frac{\partial V}{\partial \nu}\right|_{\nu^*}\frac{d\nu^*}{d\sigma}
\end{equation}
Since $\frac{\partial V}{\partial \nu}=-\int\ldots \tau u'(C)C<0$, if $\frac{d\nu^*}{d\sigma}<0$ the adjustment of the optimal policy will mitigate the effect of volatility, whereas if $\frac{d\nu^*}{d\sigma}>0$ it will accentuate it.

In order to get more specific results we have to specialize to specific forms of the utility function.

\subsection{The case $u(C)=-C^{-\gamma}$, $\gamma>0$}

If we take $u(C)=-C^{-\gamma}$ then, because of Eq. (\ref{accp}), we have $\frac{\partial V}{\partial k_0}=-\gamma V$. The behavior of accounting price is clearly related to that of the value function.

Integrals can be performed easily and we get:
\begin{equation}
V(k_0)=-\frac{(\nu k_0)^{-\gamma}}
{\delta+\gamma(\mu-\nu)-\gamma(1+\gamma)\sigma^2/2}
\end{equation}
This diverges when $\sigma\to \sigma_c$ 
from below where the critical volatility is
\begin{equation}
\sigma_c^2=2\frac{\delta+\gamma(\mu-\nu)}{\gamma(1+\gamma)}
\end{equation}

The fact that $V\to -\infty$ means that the integral which defines 
$V$ is dominated by the region $k(t)\approx 0$. This means that the economy is going bankrupt ($k=0$) after some time.

It is easy to see that $V$ decreases with $\sigma$ but the accounting 
price $p$ increases with $\sigma$. Indeed 
$2u''(C)+Cu'''(C)=\gamma(1+\gamma)^2 c^{-\gamma-2}>0$ in this case.

The wealth change is given by:
\begin{equation}
\frac{dV}{dt}=\frac{\gamma[\mu-\nu-(1+\gamma)\sigma^2/2]
(\nu k_0)^{-\gamma}}
{\delta+\gamma(\mu-\nu)-\gamma(1+\gamma)\sigma^2/2}
\end{equation}
which becomes negative if $\sigma^2>2(\mu-\nu)/(1+\gamma)$. Note that only
accounting for the change in capital stock without taking into account
the Ito term would overestimate wealth:
\begin{equation}
k_0\frac{\partial V}{\partial k_0}=\frac{\gamma(\mu-\nu)
(\nu k_0)^{-\gamma}}
{\delta+\gamma(\mu-\nu)-\gamma(1+\gamma)\sigma^2/2}
\end{equation}
This is always positive (provided $\mu>\nu$). The optimal consumption rate $\nu^*$ is, in this case
\begin{equation}
\nu^*=\frac{\delta+\gamma\mu}{1+\gamma}-\gamma\frac{\sigma^2}{2}
\end{equation}
and it decreases with $\nu$. Note that $\nu^*\to 0$ as $\sigma\to\sigma_c$.
The adjustment in the optimal consumption rate compensates the effect of volatility, but it is insufficient to change the main conclusions, i.e. $V$ still decreases with $\sigma$ and $p$ increases with it.

\subsection{The case $u(C)=C^{\beta}$, $0<\beta<1$}

When the utility function is $u(C)=C^\beta$ we have $u'(C)C=\beta u(C)$ so the behavior of the value function $V$ and of accounting prices $p=V/\beta$ is the same. In particular we find:
\begin{equation}
V(k_0)=\frac{(\nu k_0)^{\beta}}
{\delta-\beta(\mu-\nu)+\beta(1-\beta)\sigma^2/2}
\end{equation}
Concavity requires $\beta<1$. This may diverge if $\delta+\nu<\mu$ because 
capital grows very fast end the integral is dominated by the $C\to\infty$ 
region. In this case volatility smooths the divergence. 

Again $V$ decreases with $\sigma$ but now also the accounting price 
decreases with $\sigma$. Indeed 
$2u''(C)+Cu'''(C)=-\beta^2(1-\beta) c^{\beta-2}<0$ in this case.

Wealth change is given by:
\begin{equation}
\frac{dV}{dt}=\frac{\beta[\mu-\nu-(1-\beta)\sigma^2/2](\nu k_0)^{\beta}}
{\delta-\beta(\mu-\nu)+\beta(1-\beta)\sigma^2/2}
\end{equation}
again the Ito term can turn this negative if
$\sigma^2>2(\mu-\nu)/(1-\beta)$. Again neglecting the Ito term
overestimates wealth changes. 

The optimal consumption rate is
\[
\nu^*=\frac{\delta-\beta\mu}{1-\beta}+\beta\frac{\sigma^2}{2}
\]
which is clearly increasing with $\sigma$. Accounting prices, on the 
optimal policy, decrease with $\sigma$ even more strongly than with $\nu$ fixed.

\subsection{The case $u(C)=\ln C$}

In this case it is easy to check that
\begin{equation}
V(k_0)=\frac{\ln (\nu k_0)}{\delta}+\frac{\mu-\nu-\sigma^2/2}{\delta^2}
\end{equation}
and again $V$ decreases with $\sigma$ but accounting prices are independent of $\sigma$ as well as the optimal consumption rate. Now
\begin{equation}
\frac{dV}{dt}=\mu-\nu-\frac{\sigma^2}{2}
\end{equation}
and the same considerations as above apply.

\section*{Acknowledgments}

My gratitude goes to Partha Dasgupta and Karl-Goran M\"aler for many pleasant and interesting discussions on this and other topics.

\end{document}